\documentclass[%
 reprint,
superscriptaddress,
preprintnumbers,
nofootinbib,
 amsmath,amssymb,
 aps,
 prd,
floatfix,
]{revtex4-2}
\usepackage{mathtools}
\usepackage{graphicx}
\usepackage{dcolumn}
\usepackage{bm}
\usepackage[export]{adjustbox}
\usepackage{aas_macros}
\usepackage{hyperref}
\hypersetup{colorlinks = true, citecolor = blue, linkcolor = blue}

\DeclareUnicodeCharacter{2212}{-}
\begin{document}


\title{Search for gravitational waves from Scorpius X-1 with a hidden Markov model in O3 LIGO data with a corrected orbital ephemeris}

\author{Andr\'es F. Vargas}
\email{afvargas@student.unimelb.edu.au}
\author{Andrew Melatos}%
\affiliation{School of Physics, University of Melbourne, Parkville, Victoria, 3010, Australia}%
\affiliation{OzGrav-Melbourne, Australian Research Council Centre of Excellence
for Gravitational Wave Discovery, Parkville, Victoria, 3010, Australia}

\date{\today}

\begin{abstract}
Results are presented for a semi-coherent search for gravitational waves from the low-mass X-ray binary Scorpius X-1  in Observing Run 3 (O3) data from the Laser Interferometer Gravitational Wave Observatory, using an updated orbital parameter ephemeris and a hidden Markov model (HMM) to allow for spin wandering. The new orbital ephemeris corrects errors in previously published orbital measurements and implies a new search domain. This search domain does not overlap with the one used in the original Scorpius X-1 HMM O3 search. The corrected domain is approximately three times smaller by area in the $T_{\rm asc}$--$P$ plane than the original domain,  where $T_{\rm asc}$ and $P$ denote the time of passage through the ascending node and the orbital period respectively, reducing the trials factor and computing time. No evidence is found for gravitational radiation in the search band from 60 Hz to 500 Hz. Upper limits are computed for the characteristic gravitational wave strain. They are consistent with the values from the original Scorpius X-1 HMM O3 search.
\end{abstract}

\maketitle

\section{\label{sec:INTRODUCTON} INTRODUCTION}

Low-mass X-ray binaries (LMXBs) are targets of numerous gravitational wave searches \cite{AasiAbbott2015,AbbottAbbott2019a,MiddletonClearwater2020,ZhangPapa2021,TheLIGOScientificCollaborationtheVirgoCollaboration2021a,AbbottAbe2022b,AbbottAbe2022a,WhelanTenorio2023}. They are postulated to radiate continuous wave (CW) signals while in a state of rotational equilibrium, where the accretion torque balances the gravitational radiation reaction torque~\cite{PapaloizouPringle1978,Wagoner1984,Bildsten1998,Riles2013}. X-ray bright LMXBs are particularly promising targets for CW searches, because the characteristic wave strain $h_{0}$, assuming torque balance, is proportional to the square root of the X-ray flux $F_{X}$. The brightest LMXB is Scorpius X-1 (Sco X-1), whose long-term average X-ray flux is measured at $F_{X}=3.9\times 10^{-7}~{\rm erg}~{\rm cm}^{-2}{\rm s}^{-1}$~\cite{WattsKrishnan2008}.

Sco X-1 has been a regular target of CW searches with data from the Laser Interferometer Gravitational-Wave Observatory (LIGO) first (O1)~\cite{AbbottAbbott2017b,AbbottAbbott2017c,AbbottAbbott2017e}, second (O2)~\cite{AbbottAbbott2019a,ZhangPapa2021}, and third (O3)~\cite{AbbottAbbott2021b,AbbottAbe2022b,AbbottAbe2022a,WhelanTenorio2023} observing runs. Although no direct evidence of gravitational radiation has been detected, astrophysically interesting upper limits on $h_{0}$ have been obtained. For O3, a hidden Markov model (HMM) pipeline~\cite{SuvorovaClearwater2017,AbbottAbe2022b} obtained an upper limit at $95\%$ confidence of $h^{95\%} \lesssim 6 \times10^{-26}$, assuming circular polarization, across the $100$--$200~{\rm Hz}$ band. For the same frequency band, a cross-correlation (CrossCorr) pipeline~\cite{DhurandharKrishnan2008,WhelanSundaresan2015,AbbottAbe2022a} obtained $h_{0}^{95\%} \lesssim 4\times10^{-26}$, assuming the same polarization.

Searching for CW signals from Sco X-1 presents three challenges: (i) its spin frequency $f_{\star}$ (which is related to the CW frequency) is unknown, as the system does not exhibit X-ray pulsations~\cite{Riles2013,WattsKrishnan2008}; (ii) $f_{\star}$ may wander stochastically due to fluctuations in the hydromagnetic accretion torque~\cite{deKoolAnzer1993,WattsKrishnan2008}; and (iii) the signal model requires precise measurements of the orbital elements: the projected semimajor axis $a_{0}$, the time of passage through the ascending node $T_{\rm asc}$, and the orbital period $P$. The HMM and CrossCorr pipelines are both impacted by (iii). Previous CW searches for Sco X-1 have utilized orbital elements from the campaign of electromagnetic observations known as ``Precision Ephemerides for Gravitational-Wave Searches" (PEGS)~\cite{GallowayPremachandra2014,WangSteeghs2018,KillesteinMould2023} to define the astrophysically motivated parameter domain of the search. The latest update, PEGS IV~\cite{KillesteinMould2023}, offers a refined ephemeris and corrects errors in the orbital elements measured by PEGS I~\cite{GallowayPremachandra2014} and III~\cite{WangSteeghs2018}\footnote{PEGS II~\cite{PremachandraGalloway2016} targeted Cygnus X-2 instead of Sco X-1.}. Notably, the $T_{\rm asc}$--$P$ domain from PEGS III, used in the O2~\cite{AbbottAbbott2019a,ZhangPapa2021} and O3 analyses~\cite{AbbottAbe2022b,AbbottAbe2022a}, does not overlap with the domain from PEGS IV. A revised Sco X-1 HMM search, using O3 LIGO data~\cite{AbbottAbe2023} and the new PEGS IV ephemeris, is necessary and forms the subject of this paper.

The outline of the paper is as follows. Section~\ref{sec:OrbEphmScoX1} discusses briefly the differences in the measurements and results between PEGS III and PEGS IV. Section~\ref{sec:SearchImplementation} summarizes the HMM search pipeline and the parameter domain. Sections~\ref{sec:ReanalysisO3} and~\ref{sec:UL} present the re-analysis output and revised upper limits, respectively. Section~\ref{sec:Conclusions} presents the conclusion. 

\section{Orbital elements}
\label{sec:OrbEphmScoX1}

An LMXB emits a quasimonochromatic CW signal~\cite{MukherjeeMessenger2018,TenorioKeitel2021}, which is Doppler-modulated by the relative motion of the source and the detector. The Doppler modulation depends on the sky position (right ascension $\alpha$, declination $\delta$) and orbital elements of the binary. Accurate electromagnetic measurements of $\alpha,\delta,a_{0},T_{\rm asc}$, and $P$ reduce the computational cost and trials factor of the search, thereby increasing its sensitivity at a fixed false alarm probability.

For Sco X-1, the sky position is known with microarcsecond precision~\cite{BradshawFomalont1999,AbbottAbbott2017b}. The orbital eccentricity, $e$, satisfies $e \leq 0.0132$~\cite{KillesteinMould2023} and is set to zero by assumption in this paper and previous searches~\cite{AbbottAbbott2019a,AbbottAbe2022b}. The other orbital elements, i.e. $a_{0}$, $T_{\rm asc}$, and $P$, are harder to measure because of the high accretion rate. The PEGS program tracks the orbital motion via the emission lines generated by Bowen fluorescence of the irradiated donor star~\cite{SteeghsCasares2002,WangSteeghs2018,KillesteinMould2023}. The orbital elements are inferred by fitting a Keplerian orbit ($e=0$; see Equation~(1) in Ref.~\cite{WangSteeghs2018}) to the modulated radial velocities of the Bowen emission lines. The semimajor axis $a_0$ is related to the orbital speed $K_1$, inferred from the Keplerian fit, by $a_0=K_{1}P/(2\pi)$. PEGS III obtains $40~{\rm km\,s}^{-1} \leq K_{1} \leq 90~{\rm km\,s}^{-1}$, or equivalently $1.45~{\rm lt}$-s $\leq a_{0} \leq 3.25~{\rm lt}$-s~\cite{WangSteeghs2018}. The orbital phase, determined by $T_{\rm asc}$ and $P$, is dominated by uncertainties related to time-referencing, and the covariance ${\rm cov}(P,T_{\rm asc})$. The former refers to possible errors in the barycentric time, for example, while the latter refers to the covariance of the parameters inherent to the Keplerian fit.

The PEGS IV ephemeris~\cite{KillesteinMould2023} improves on previously published PEGS updates~\cite{GallowayPremachandra2014,WangSteeghs2018,KillesteinMould2023}. PEGS IV is derived from over $20$ yr of high-quality spectroscopic observations from the Visual Echelle Spectrograph mounted on the Very Large Telescope. The data, which consist of optical spectra of Bowen emission lines, are analyzed with a new Bayesian approach which supersedes the least-square fit in PEGS III, accounts for observational systematics, and maximizes the precision of the uncertainty estimates~\cite{KillesteinMould2023}. The Bayesian re-analysis reveals two calibration errors in previous ephemerides. First, PEGS I and III do not correct for the mid-exposure time, i.e. the mid time of the various spectroscopic observations. Moreover, they do not adjust for the Solar System heliocentre, which induces a systematic $T_{\rm asc}$ error of $\approx 0.6~{\rm k s}$. Second, the covariances between orbital elements are underestimated by two orders of magnitude in the Keplerian fit~\cite{KillesteinMould2023}. 

Table~\ref{tab:PEGSIIIvsIV} summarizes the values inferred by PEGS III and IV for $a_{0},T_{\rm asc}$ and $P$. The discrepancies between $T_{\rm asc,ref}$ and $P$ in the second and third columns are apparent. Henceforth we define $T_{\rm asc,ref}$ as the time of ascension corresponding to November 21 23:16:49 UTC 2010 for PEGS III, and March 06 15:07:40 UTC 2014 for PEGS IV. 

\begin{figure}
\centering
    \includegraphics[scale=0.42]{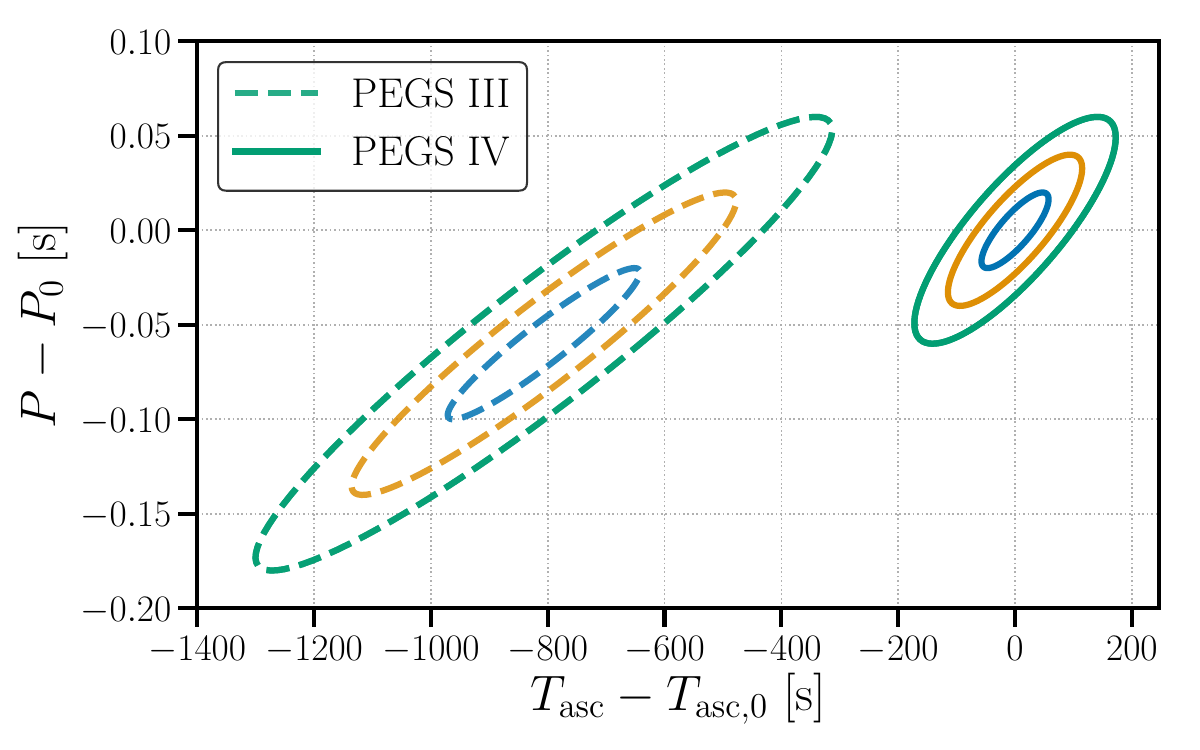}
    \caption{Forward propagated PEGS III (dashed contours) and PEGS IV (solid contours) uncertainty ellipses corresponding to $1\sigma$ (blue line), $2\sigma$ (yellow line), and $3\sigma$ (green line), for $T_{\rm asc}$ and $P$. The values of $T_{\rm asc,ref}$ (see Table~\ref{tab:PEGSIIIvsIV}) are propagated by $N_{\rm orb}=2352$ and $N_{\rm orb}=3877$ to generate the solid and dashed curves, respectively. $P_{0}$  represents the central PEGS IV value of $P$, i.e. $P_{0}=68\,023.92~{\rm s}$, while $T_{\rm asc,0}$ represents the central PEGS IV value of $T_{\rm asc,ref}$ propagated to the start of O3 (dated $T_{\rm O3,0}=1\,238\,166\,483$ GPS time), i.e. $T_{\rm asc,0}=1\,238\,145\,935.84~{\rm s}$. The HMM O3 search in this paper covers the region enclosed by the solid green contour.} 
    \label{fig:propagation-Ellipses-PEGS}
\end{figure}

In general, CW searches propagate the value of $T_{\rm asc,ref}$ to an equivalent time of ascension at a later epoch near or within the relevant LIGO observing run. The propagated value is given by $T_{\rm asc,0}=T_{\rm asc,ref}+N_{\rm orb}P$, independent of $a_{0}$, where $N_{\rm orb}$ is an integer number of orbits. For LIGO O3 data, which starts at $T_{\rm O3,0}=1\,238\,166\,483$ GPS time, $T_{\rm asc,ref}$ is propagated forward by $N_{\rm orb}=3877$ and $N_{\rm orb}=2352$, when using the PEGS III and PEGS IV ephemerides, respectively. The original uncertainties for $T_{\rm asc,ref}$ and $P$ are also propagated following the same recipe (see Section~\ref{subsec_II:scoX1params_and_tmplts}). Fig.~\ref{fig:propagation-Ellipses-PEGS} shows the result of propagating $T_{\rm asc,ref}$ and its uncertainties for the PEGS III (dotted contours) and IV (solid contours) ephemerides. The $T_{\rm asc}$--$P$ domains covered by the two ephemerides do not overlap at the $3\sigma$ level, with a difference of $\approx 0.8~{\rm ks}$ in $T_{\rm asc}$. The original O2 and O3 searches~\cite{AbbottAbbott2019a,ZhangPapa2021,AbbottAbe2022b,AbbottAbe2022a} covered the PEGS III domain and therefore looked in the wrong place. We rectify this error for the HMM search in this paper, just as Ref.~\cite{WhelanTenorio2023} rectifies the error for the CrossCorr search. 

\begin{table}
\begin{ruledtabular}
\caption{\label{tab:PEGSIIIvsIV} Measured orbital elements, for the PEGS III and IV ephemerides. The $T_{\rm asc, ref}$ and $P$ uncertainties correspond to $\pm 1\sigma$. The quoted values are taken from the cited references.}
\begin{tabular}{ccc}
Parameter & PEGS III~\cite{WangSteeghs2018} & PEGS IV~\cite{KillesteinMould2023} \\ \hline 
$a_{0}$ (lt-s) & [1.45,3.25] & [1.45,3.25] \\
$T_{\rm asc,ref}$ (GPS s) & $974433630\pm50$  & $1078153676\pm33$   \\
$P$ (s) & $68023.86\pm0.04$ &  $68023.92\pm0.02$  \\
\end{tabular}
\end{ruledtabular}
\end{table}

\section{Search Implementation}
\label{sec:SearchImplementation}

In this section we discuss the practical details of the re-analysis. Section~\ref{subsec_II:HMMsearch} briefly presents the HMM search algorithm. Section~\ref{subsec_II:scoX1params_and_tmplts} defines the search domain and the orbital template grid. Section~\ref{subsec_II:thresholds} updates the detection thresholds used in the previous analysis. Section~\ref{subsec_II:Vetoes} reviews briefly the vetoes used in the re-analysis.

\subsection{HMM}
\label{subsec_II:HMMsearch}

A HMM is characterized by a hidden state variable $q(t)$, which takes the discrete values $\{q_{1},...,q_{N_{Q}}\}$, and an observable state $o(t)$, which takes the discrete values $\{o_{1},...,o_{q_{N_{Q}}}\}$. The HMM tracks a stochastic process which jumps between hidden state values at discrete epochs $\{t_{0},...,t_{N_{T}}\}$. In this CW application, $f_{\star}(t)$ is mapped onto $q(t)$, to track the wandering CW emission frequency from one time step to the next.  The probability to jump from $q_{i}$ at $t_{n}$ to $q_{j}$ at $t_{n+1}$ is given by the transition matrix $A_{q_{j}q_{i}}$~\cite{AbbottAbe2022b}. The transitions are modeled as an unbiased random walk, in which $f_{\star}(t)$ jumps by $-1,0$, or $1$ frequency bins, of width $\Delta f_{\rm drift}$, with equal probability at each epoch. 

The Fourier transform of the time-domain O3 data maps onto the observable states. The O3 data, with duration $T_{\rm obs}$, is divided into $N_{T}$ segments of duration $T_{\rm drift}$, viz. $N_{T}={\rm floor}(T_{\rm obs}/T_{\rm drift})$. We discuss the choice of $T_{\rm drift}$ for this search in Section~\ref{subsec_II:scoX1params_and_tmplts}. The emission probability matrix $L_{o_{j}q_{i}}$ relates the observed data between two consecutive epochs, $\{o(t') \vert~t_{j} \leq t' \leq t_{j}+T_{\rm drift}\}$, to the occupied hidden state $q(t_{i})$.  We set $L_{o_{j}q_{i}}$ proportional to the ${\cal J}$-statistic~\cite{SuvorovaClearwater2017,AbbottAbbott2019a}, a frequency domain estimator which ingests the search frequency $f_{0}$ and the orbital elements $(a_{0},T_{\rm asc},P)$ and accounts for the orbital Doppler shift of the CW carrier frequency. 

The probability that the hidden state follows some path $Q=\{q(t_{0}),...,q(t_{N_{T}})\}$ given the observed data $O=\{o(t_{0}),...,o(t_{N_{T}})\}$ is given by

\begin{equation}
    P(Q\vert O) = \Pi_{q(t_{0})} \prod_{i=1}^{N_{T}} L_{o(t_{i})q(t_{i})} A_{q(t_{i})q(t_{i-1})},
    \label{Eq:P_Q_gvn_O}
\end{equation}

\noindent where $\Pi_{q(t_{0})}$ is the prior probability (here uniform, with $\Pi_{q(t_{0})}=1/N_{Q}$) of being in some initial state $q(t_{0})$ at time $t_{0}$. We use the Viterbi algorithm~\cite{AViterbi1967} to calculate the path $Q^{*}$ that maximizes $P(Q\vert O)$ in Equation~(\ref{Eq:P_Q_gvn_O}). We use the log-likelihood of the most likely path ${\cal L}=\ln P(Q\vert O)$, output by the Viterbi algorithm, as the detection statistic.

The HMM scheme used in this paper is identical to the one used in the original O3 search~\cite{AbbottAbe2022b} and Refs.~\cite{SuvorovaClearwater2017,AbbottAbbott2019a,MiddletonClearwater2020,TheLIGOScientificCollaborationtheVirgoCollaboration2021a,VargasMelatos2023}. For a description of the workflow, see Section~III C and the flowchart in Fig. 2 in Ref.~\cite{AbbottAbe2022b}.

\subsection{Revised signal templates based on PEGS IV ephemeris}
\label{subsec_II:scoX1params_and_tmplts}
\begin{table*}
\caption{\label{tab:ScoX1Params} Search parameters and their ranges. The column headed ``EM data" records the availability of electromagnetic measurements drawn from the references in the last column. As written here $T_{\rm asc}$ and $P$, plus their uncertainties, define a rectangular parameter domain; values without uncertainties are treated as a single template. In the text generally the central values are denoted with the subscript $0$, e.g. $P_{0}$, or and overbar, e.g. $\bar{a}_{0}$. The time of ascension $T_{\rm asc}$ stands for the  value in \cite{KillesteinMould2023} propagated up to the start of O3, as described in Section \ref{sec:OrbEphmScoX1}.  We only analyze the templates that are enclosed within the propagated $T_{\rm asc}$-$P$ prior ellipses, i.e. the green solid ellipses in Fig.~\ref{fig:propagation-Ellipses-PEGS}; see Section \ref{subsec_II:scoX1params_and_tmplts} for details.}
\begin{ruledtabular}
\begin{tabular}{ccccc}
 Parameter & Symbol & Search range & EM data & Reference\\ \hline
 Right ascension& $\alpha$ & $16\,\text{h}\,19\,\text{m}\,55.0850\,\text{s}$ & Y &\cite{BradshawFomalont1999} \\
 Declination& $\delta$ & $-15^{\circ}38'24.9''$& Y & \cite{BradshawFomalont1999}\\
 Orbital inclination angle & $\iota$ & $44\,\pm6^{\circ}$ & Y & \cite{FomalontGeldzahler2001} \\
 Projected semi-major axis & $a_{0}$ & $2.35\pm0.3\,\text{lt-s}$& Y & \cite{WangSteeghs2018}\\
 Orbital period& $P$ & $68\,023.92\,\pm\,0.02\,\text{s}$& Y & \cite{KillesteinMould2023}\\
 GPS time of ascension & $T_{\rm asc}$ & $1\,238\,145\,935.84\pm\,57\,\text{s}$ & Y & \cite{KillesteinMould2023} \\
 Frequency& $f$ & $60\,-\, 500\,\text{Hz}$ & N & ... \\
\end{tabular}
\end{ruledtabular}
\end{table*}

The signal templates searched in the PEGS IV re-analysis are not the same as those searched in the original PEGS III analysis. Specifically, the template locations change in $T_{\rm asc}$ and $P$ but not in $f_0$, $a_0$, and sky position.

We set $T_{\rm drift}=10~{\rm days}$, $\Delta f_{\rm drift} = 1/(2T_{\rm drift})=5.787037\times10^{-7}~{\rm Hz}$ and $N_{T}=36$. The choice of $T_{\rm drift}$ is motivated astrophysically and historically. It is the characteristic timescale of the random walk in $f_{\star}(t)$ inferred from the accretion-driven fluctuations in the X-ray flux of Sco X-1~\cite{AasiAbbott2015,MukherjeeMessenger2018,MessengerBulten2015}. It also matches the previously published Sco X-1 searches, and hence enables direct comparison between them~\cite{AasiAbbott2015,AbbottAbbott2017c,AbbottAbbott2017b,AbbottAbbott2019a,AbbottAbe2022b}.

The band to be searched, $60$--$500~{\rm Hz}$, is identical to the original search. It is divided into sub-bands, each of width $\Delta f_{\rm sub}=0.608148~{\rm Hz}$ and containing $N_{f}=2^{20}$ frequency bins. The total number of sub-bands is $N_{\rm sub}={\rm ceil}[(500-60)~{\rm Hz}/\Delta f_{\rm sub}]=725$. 

The ${\cal J}$-statistic depends on the orbital parameters $(a_{0},T_{\rm asc},P)$ in addition to the location of the source, described by the right ascension $\alpha$ and declination $\delta$. These parameters have been measured electromagnetically for Sco X-1~\cite{BradshawFomalont1999,FomalontGeldzahler2001,WangSteeghs2018,KillesteinMould2023}. In this re-analysis the template grid remains unchanged from the original analysis except in the $T_{\rm asc}$-$P$ plane~\cite{AbbottAbe2022b}. In particular, $\alpha$, $\delta$, the orbital inclination angle $\iota$, and the $a_{0}$ domain, are unchanged. Following Section~\ref{sec:OrbEphmScoX1}, the $a_{0}$ domain is defined by the range $1.45 \leq a_{0} /(\text{lt-s}) \leq 3.25$. Equivalently, we express this range as $\bar{a}_{0}\pm3\sigma_{a_{0}}$, with $\bar{a}_{0}=2.35$ lt-s and $\sigma_{a_{0}}=0.3$ lt-s. The grid spacing for $a_{0}$ is discussed in the next paragraph. As explained in Section~\ref{sec:OrbEphmScoX1}, we propagate forward the reference value $T_{\rm asc,ref}$ (see Table~\ref{tab:PEGSIIIvsIV}) to the start of O3, $T_{\rm O3,0}$, by adding an integer number of orbits $N_{\rm orb}$. This yields a central value $T_{\rm asc,0}=1\,238\,145\,935.84~{\rm s}$. The uncertainties on $T_{\rm asc,ref}$ are also propagated to the start of O3 via $\sigma_{T_{\rm asc}}=[ \sigma^{2}_{T_{\rm asc,ref}}+(N_{\rm orb}\sigma_{P})^{2}]^{1/2}$. The latter equation yields $\sigma_{T_{\rm asc}}=57~{\rm s}$ for $\sigma_{T_{\rm asc,ref}}=33~{\rm s}$ and $\sigma_{P}=0.02~{\rm s}$. We summarize the search domain in Table~\ref{tab:ScoX1Params}.

We cover the orbital parameter domain with a rectangular grid spanning $(\bar{a}_{0}\pm3\sigma_{a_{0}},T_{\rm asc,0}\pm3\sigma_{T_{\rm asc}}, P_{0}\pm3\sigma_{P})$. We use Equation~(71) of Ref.~\cite{LeaciPrix2015} to set the spacing of the grid and the number of grid points $N_{a_{0}},N_{T_{\rm asc}}$, and $N_{P}$. For this search we adopt a maximum mismatch of $\mu_{\rm max}=0.1$, as for the original PEGS III search. Table~\ref{tab:guided_freqs_and_Ns} presents the number of grid points for several selected sub-bands using the PEGS IV ephemeris. As with the original search, we only analyze templates within the $3\sigma$ uncertainty ellipses (green solid contour in Fig.~\ref{fig:propagation-Ellipses-PEGS}). Covering the updated $(T_{\rm asc,0}\pm3\sigma_{T_{\rm asc}}, P_{0}\pm3\sigma_{P})$ domain requires $\approx 3$ and $\approx 2$ times fewer $N_{T_{\rm asc}}$ and $N_{P}$ grid points, respectively, when averaged across all sub-bands. For instance, the $160~{\rm Hz}$ sub-band in the original search uses $N_{T_{\rm asc}}=394$ and $N_{P}=12$ (see Table II in Ref.~\cite{AbbottAbe2022b}), while this search uses $N_{T_{\rm asc}}=117$ and $N_{P}=6$. Overall we analyze between  $ 1.03\times10^{5}$ and $3.81\times10^{7}$ templates per sub-band, compared to between $5.71\times10^{5}$ and $2.88\times10^{8}$ templates for the PEGS III uncertainty ellipses (dashed contours in Fig.~\ref{fig:propagation-Ellipses-PEGS}). 

Besides a rectangular grid, as explained above, there are other alternatives for covering the $T_{\rm asc}$--$P$ domain with templates. For example, using sheared period coordinates~\cite{WagnerWhelan2022} or lattice-tilling template banks~\cite{MukherjeePrix2023}. 

\begin{table}
\caption{Selected sub-bands and corresponding $N_{a_{0}}, N_{T_{\rm asc}}, N_{P}$ template counts needed to cover the $(\bar{a}_{0}\pm3\sigma_{a_{0}},T_{\rm asc,0}\pm3\sigma_{T_{\rm asc}}, P_{0}\pm3\sigma_{P})$ domain, inferred from the parameter space metric with $\mu_{\rm max}=0.1$~\cite{LeaciPrix2015}. The orbital templates $N_{a_{0}}N_{T_{\rm asc}}N_{P}$ per sub-band vary by a factor of $\approx 385$ across the full band $60$--$500~{\rm Hz}$. Table II in Ref.~\cite{AbbottAbe2022b} indicates that $N_{T_{\rm asc}}$ and $N_P$ for PEGS III are $\approx 3$ and $\approx2$ times lower than for PEGS IV respectively.} 
\label{tab:guided_freqs_and_Ns} 
\begin{ruledtabular}
\begin{tabular}{cccc}
Sub-band (Hz) & $N_{a_{0}}$ & $N_{T_{\rm asc}}$ & $N_{P}$ \\ \hline
$60$ & $767$ & $45$ & $3$ \\
$160$ & $2031$ & $117$ & $6$ \\
$260$ & $3296$ & $190$ & $9$ \\
$360$ & $4560$ & $263$ & $13$ \\
$500$ & $6331$ & $364$ & $17$ \\
\end{tabular}
\end{ruledtabular}
\end{table}

\subsection{Updated detection threshold}
\label{subsec_II:thresholds}

A sub-band is flagged as a candidate when the optimal path with highest log-likelihood, denoted ${\rm max}({\cal L})$, exceeds a threshold ${\cal L}_{\rm th}$, corresponding to a user-selected, fixed, false alarm probability. The process to obtain ${\cal L}_{\rm th}$ in each sub-band, via Monte-Carlo simulations, is described in Section III D of Ref.~\cite{AbbottAbe2022b} and Appendix A in Ref.~\cite{TheLIGOScientificCollaborationtheVirgoCollaboration2021a}. 

In general, ${\cal L}_{\rm th}$ depends on the trials factor associated with each sub-band, i.e. $N_{\rm tot}=N_{f}N_{a_{0}}N_{\rm bin}$, where $N_{\rm bin}$ is the number of templates required to cover the $T_{\rm asc}$--$P$ domain per sub-band. The false alarm probability $\alpha_{N_{\rm tot}}$ per sub-band is given by

\begin{equation}
    \alpha_{N_{\rm tot}}=1-(1-\alpha)^{N_{\rm tot}},
    \label{Eq:FAP}
\end{equation}

\noindent where $\alpha$ is the false alarm probability in a single terminating frequency bin per orbital template. Here we follow previous Sco X-1 HMM searches~\cite{AbbottAbbott2017c,AbbottAbbott2019a,AbbottAbe2022b} and adopt $\alpha_{N_{\rm tot}}=0.01$ to avoid excessive follow-up of candidates and facilitate the comparison between searches. 

Given that $N_{\rm bin}$ is different in this search, we update our thresholds accordingly. Based on the new values of $N_{\rm tot}$ per sub-band from Table~\ref{tab:guided_freqs_and_Ns}, we obtain ${\cal L}({\rm PEGS~III}) - {\cal L}({\rm PEGS~IV}) \approx 5, 6$, and $8$ for the minimum, average, and maximum across all sub-bands. The difference varies monotonically across the search band.

\subsection{Vetoes}
\label{subsec_II:Vetoes}

All sub-bands with ${\rm max}({\cal L})>{\cal L}_{\rm th}$ are subjected to a hierarchy of vetoes to distinguish between a non-Gaussian instrumental artifact and a possible astrophysical signal. In this paper, we follow the original O3 HMM search~\cite{AbbottAbe2022b} and apply the known lines veto and the single interferometer (IFO) veto. The known lines veto eliminates any candidate overlapping with a narrow-band noise artifact listed in Ref.~\cite{O3lines}. The single IFO veto searches for the candidate in both detectors separately to eliminate any candidate caused by a noise artifact present in one of the detectors only. These vetoes have been used extensively in previous HMM searches~\cite{AbbottAbbott2017c,AbbottAbbott2019a,MiddletonClearwater2020,TheLIGOScientificCollaborationtheVirgoCollaboration2021a,AbbottAbe2022b}. They are defined and justified in detail in the latter references.

\section{Re-analysis of LIGO O3 data}
\label{sec:ReanalysisO3}

The re-analysis described in Section~\ref{sec:SearchImplementation} yields $23$ candidates that satisfy ${\rm max}({\cal L})>{\cal L}_{\rm th}$. From these candidates, the known lines veto eliminates $21$, and the single IFO veto eliminates two.

\begin{figure*}[ht]
    \centering
    \includegraphics[scale=0.375, left]{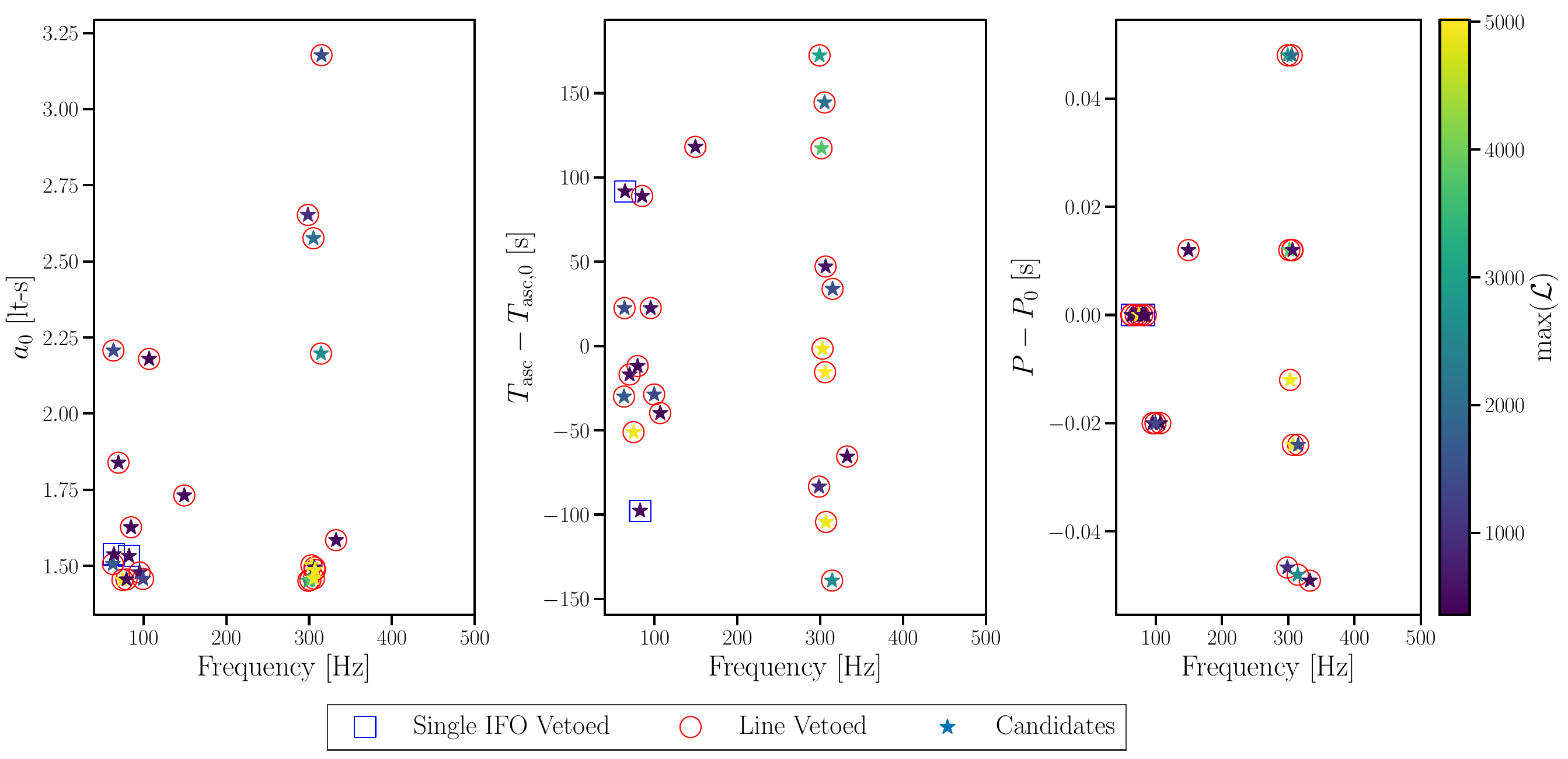}
    \caption{Candidates (stars) plotted as a function of their terminating frequency bin $q^{*}(t_{N_{T}})$ (horizontal axis). The vertical axes feature the orbital parameters: $a_{0}$ (left panel), offset from the central time of ascension $T_{\rm asc}-T_{\rm asc, 0}$ (middle panel), and offset from the central period $P-P_{0}$ (right panel). The colour scale corresponds to $\rm max(\mathcal{L})$. Candidates decorated with a red circle or blue square are eliminated by the known lines or single IFO veto, respectively.}
    \label{fig:CandidatesperSub}
\end{figure*}

We plot the results of the re-analysis and the veto procedure in Fig.~\ref{fig:CandidatesperSub}. The horizontal axis, for all panels, correspond to the terminating frequency bin of the optimal path, $q^{*}(t_{N_{T}})$. The vertical axes correspond to the orbital elements: $a_{0}$ (left panel), $T_{\rm asc}-T_{\rm asc,0}$ (central panel), and $P-P_{0}$ (right panel). Candidates eliminated by the known lines veto and the single IFO veto are marked with a red circle and a blue square, respectively.

This re-analysis also gives us an opportunity to check whether the vetoed candidate sub-bands depend on the orbital ephemerides, i.e. see whether sub-bands consistently get flagged as candidates, regardless of the ephemerides we use. With the exception of the $106.78~{\rm Hz}$ sub-band, all other candidate sub-bands were previously flagged in the original O3 analysis. Among these shared sub-bands, those commencing at $64.30~{\rm Hz}$ and $82.51~{\rm Hz}$ were vetoed in both analyses by the single IFO veto, while the rest are vetoed by the known lines veto.

\begin{table*}[tbp]
\caption{\label{tab:Candidates} Candidates yielded by the O3 re-analysis. The first and second columns correspond to the starting frequency of the sub-band containing the candidate, and the log-likelihood of the candidate, respectively. The third and fourth columns record the outcome of the two vetoes described in Section \ref{subsec_II:Vetoes}. Candidates are marked with \checkmark if they pass the veto or \textbf{X} if they do not. H and L correspond to the ${\rm max}({\cal L})$ obtained using Hanford-only and Livingston-only data.}
\begin{ruledtabular}
\begin{tabular}{cccc}
Sub-band $(\text{Hz})$ & $\rm max(\mathcal{L})/10^{3}$ & Known lines veto & Single IFO veto \\ \hline 
63.09 &	1.65 &	\textbf{X} &	 ...  \\ 
63.70 &	1.41 &	\textbf{X} &	...  	\\
64.30 &	0.38 &	\checkmark &	\textbf{X}   \\ 
69.76 &	0.52 &	\textbf{X} &	...  \\ 
74.62 &	4.93 &	\textbf{X} &	...   \\
79.47 &	0.43 &	\textbf{X} &	...  \\
82.51 &	0.49 &	\checkmark & \textbf{X}  	\\
84.93 &	0.40 &	\textbf{X} &	...    \\
95.25 & 0.57 &	\textbf{X} &	...  \\
99.50 &	1.34 &	\textbf{X} &	... \\
106.78 & 0.36&	\textbf{X} &	...   \\
149.26 & 0.46&	\textbf{X} &	...  \\
298.53 & 0.95&	\textbf{X} &	...  \\
299.14 & 3.02&	\textbf{X} &	...  \\
301.57 & 3.73&	\textbf{X} &	...  \\
302.78 & 4.91&	\textbf{X} &	...  \\
305.21 & 2.05&	\textbf{X} & ...  \\
305.81 & 5.01&	\textbf{X} &	...  \\
306.42 & 0.61&	\textbf{X} &	... \\
307.03 & 4.94&	\textbf{X} &	...  \\
314.31 & 2.63&	\textbf{X} & ... \\
314.92 & 1.47&	\textbf{X} & ...  \\
332.51 & 0.43&	\textbf{X} & ...  \\ \hline 
Total: 23 &  &	 &	  \\
\end{tabular}
\end{ruledtabular}
\end{table*}

\section{Frequentist upper limits}
\label{sec:UL}

Historically, HMM Sco X-1 searches~\cite{AbbottAbbott2017c,AbbottAbbott2019a,AbbottAbe2022b} use non-candidate sub-bands to set upper limits on the gravitational wave strain detectable at $95\%$ confidence, $h_{0}^{95\%}$. It is important to check by how much --- if at all --- $h_0^{95\%}$ changes when moving from the PEGS III to the PEGS IV ephemerides.

Given the different detection thresholds in the PEGS III and PEGS IV ephemerides (see Section~\ref{subsec_II:thresholds}), we check as a precaution for any changes in $h_{0}^{95\%}$ from the original O3 analysis, as presented in Fig 4. of~Ref.~\cite{AbbottAbe2022b}. To this end, we apply the same frequentist upper limit procedure used in the PEGS III analysis to selected sub-bands starting at $61.27~{\rm Hz}, 174.14~{\rm Hz}, 260.31~{\rm Hz}, 360.43~{\rm Hz}$, and $500~{\rm Hz}$, spaced across the entire search band. Briefly, the upper limit procedure consists of injecting a Sco X-1-like signal, with parameters $\{h_{0},a_{0},T_{\rm asc},P\}_{\rm inj}$ and $\vert \cos \iota \vert=1$, into a sub-band. The value of $h_{0}$ is progressively reduced, while holding $\{a_{0},T_{\rm asc},P\}_{\rm inj}$ constant, until the difference in $h_{0}$ between the last detected and non-detected signals is $\delta h_{0}\leq1\times10^{-27}$. Section V A of Ref.~\cite{AbbottAbe2022b} expands on the upper limit procedure.

We find that the re-analysis and the original O3 search exhibit a similar sensitivity. The average difference, across the selected sub-bands, between the PEGS III and IV upper limits is $\langle \Delta h_{0}^{95\%} \rangle =\langle h_{0}^{95\%}({\rm PEGS~III})-h_{0}^{95\%}({\rm PEGS~IV}) \rangle = 1\times10^{-27}=\delta h_{0}$. The maximum difference between $h_{0}^{95\%}({\rm PEGS~III})$ and $h_{0}^{95\%}({\rm PEGS~IV})$ is ${\rm max}(\Delta h^{95\%}_{0})\approx 3\times10^{-27}$ for the $360.43~{\rm Hz}$ sub-band, while the minimum difference is ${\rm min}(\Delta h^{95\%}_{0})\approx -1\times10^{-27}$ for the $260.31~{\rm Hz}$ sub-band.

Given that $\langle \Delta h_{0}^{95\%} \rangle \leq \delta h_{0}$, we conclude the PEGS III upper limits remain valid and unchanged for this re-analysis. In general, as the $\Delta h_{0}^{95\%}$ value suggests, the difference in detection threshold due to the revised $N_{P}$ and $N_{T_{\rm asc}}$ template numbers yields $h_{0}^{95\%}$ values marginally lower across the frequency range when compared to the PEGS III values. Likewise, the CrossCorr PEGS IV search does not update the original O3 search upper limits, depicted in Fig. 6 of Ref.~\cite{AbbottAbe2022a}, as the re-analysis sensitivity matches the original O3 search~\cite{WhelanTenorio2023}.

\section{Conclusions}
\label{sec:Conclusions}

In this paper we re-analyze the LIGO O3 data for CW signals from Sco X-1, using a HMM pipeline in tandem with the corrected and refined PEGS IV orbital ephemeris~\cite{KillesteinMould2023}. The revised search closely follows the original HMM workflow using the PEGS III ephemeris~\cite{AbbottAbe2022b}, with identical search implementation, vetoes, and upper limits procedure. No candidate survives the hierarchy of vetoes. The upper limits procedure yields $h_{0}^{95\%}$ values consistent with those derived from the search using the PEGS III ephemeris. Consequently, the $h_{0}^{95\%}$ values presented in Ref.~\cite{AbbottAbe2022b} remain valid and unaltered, e.g. we have $h_{0}^{95\%} \leq 6\times10^{-26}$, assuming circular polarization, across the $100$--$200~{\rm Hz}$ band.  The re-analysis complements the results of the CrossCorr search using the PEGS IV ephemeris~\cite{WhelanTenorio2023}, which assumes a signal model with a constant $f_{\star}$ throughout the observation. 

The upcoming fourth LIGO-Virgo-KAGRA (LVK) collaboration observing run (O4) will offer a renewed opportunity to search for Sco X-1, taking advantage of the improved sensitivity of the detectors and the improved precision and accuracy of the PEGS IV ephemeris.

\section{Acknowledgements}

The authors thank J. B. Carlin, J. T. Whelan, D. Keitel and the members of the LVK continuous waves group for helpful suggestions which improved the manuscript. This research was supported by the Australian Research Council Centre of Excellence for Gravitational Wave Discovery (OzGrav), grant number CE170100004. This work used computational resources of the OzSTAR national facility at Swinburne University of Technology. OzSTAR is funded by Swinburne University of Technology and also the National Collaborative Research Infrastructure Strategy (NCRIS). This material is based upon work supported by NSF's LIGO Laboratory which is a major facility fully funded by the National Science Foundation.

This research has made use of data or software obtained from the Gravitational Wave Open Science Center\footnote{\href{https://gwosc.org/}{https://gwosc.org/}}, a service of the LIGO Scientific Collaboration, the Virgo Collaboration, and KAGRA. This material is based upon work supported by NSF's LIGO Laboratory which is a major facility fully funded by the National Science Foundation, as well as the Science and Technology Facilities Council (STFC) of the United Kingdom, the Max-Planck-Society (MPS), and the State of Niedersachsen/Germany for support of the construction of Advanced LIGO and construction and operation of the GEO600 detector. Additional support for Advanced LIGO was provided by the Australian Research Council. Virgo is funded, through the European Gravitational Observatory (EGO), by the French Centre National de Recherche Scientifique (CNRS), the Italian Istituto Nazionale di Fisica Nucleare (INFN) and the Dutch Nikhef, with contributions by institutions from Belgium, Germany, Greece, Hungary, Ireland, Japan, Monaco, Poland, Portugal, Spain. KAGRA is supported by Ministry of Education, Culture, Sports, Science and Technology (MEXT), Japan Society for the Promotion of Science (JSPS) in Japan; National Research Foundation (NRF) and Ministry of Science and ICT (MSIT) in Korea; Academia Sinica (AS) and National Science and Technology Council (NSTC) in Taiwan.

This work has been assigned LIGO document number P2300322.

\bibliography{only_ads_references_II,only_non_ads_references}

\end{document}